\documentclass[conference]{IEEEtran}
\IEEEoverridecommandlockouts
\usepackage{cite}
\usepackage{amsmath,amssymb,amsfonts}
\usepackage{algorithmic}
\usepackage{graphicx}
\usepackage{textcomp}
\usepackage{bbm}
\usepackage{multirow}
\usepackage{hhline}
\usepackage{xcolor}

\def\BibTeX{{\rm B\kern-.05em{\sc i\kern-.025em b}\kern-.08em
    T\kern-.1667em\lower.7ex\hbox{E}\kern-.125emX}}
\usepackage{etoolbox}
\patchcmd{\thebibliography}{\chapter*}{\section*}{}{}
\begin{document}

\title{Sequence-guided protein structure determination  using graph convolutional and recurrent networks}

\author{\IEEEauthorblockN{Po-Nan Li}
\IEEEauthorblockA{\textit{Dept. of Electrical Eng.}\\
\textit{Stanford University}\\
Stanford, CA, USA \\
liponan@stanford.edu}
\and
\IEEEauthorblockN{Saulo H. P. de Oliveira}
\IEEEauthorblockA{\textit{Division of Biosciences} \\
\textit{SLAC National Accelerator Laboratory}\\
Menlo Park, CA, USA \\
oliveira@slac.stanford.edu}
\and
\IEEEauthorblockN{Soichi Wakatsuki}
\IEEEauthorblockA{\textit{Dept. of Structural Biology}\\
\textit{Stanford University}\\
Stanford, CA, USA \\
soichi.wakatsuki@stanford.edu}
\and
\IEEEauthorblockN{Henry van den Bedem}
\IEEEauthorblockA{\textit{Atomwise, Inc.}\\
San Francisco, CA, USA \\
vdbedem@atomwise.com}
}

\maketitle

\begin{abstract}
Single particle, cryogenic electron microscopy (cryo-EM) experiments now routinely produce high-resolution data for large proteins and their complexes. 
Building an atomic model into a cryo-EM density map is challenging, particularly when no structure for the target protein is known \textit{a priori}. 
Existing protocols for this type of task often rely on significant human intervention and can take hours to many days to produce an output. 
Here, we present a fully automated, template-free model building approach that is based entirely on neural networks. 
We use a graph convolutional network (GCN) to generate an embedding from a set of rotamer-based amino acid identities and candidate 3-dimensional C$\alpha$ locations. 
Starting from this embedding, we use a bidirectional long short-term memory (LSTM) module to order and label the candidate identities and atomic locations consistent with the input protein sequence to obtain a structural model. 
Our approach paves the way for determining protein structures from cryo-EM densities at a fraction of the time of existing approaches and without the need for human intervention. 
\end{abstract}


\section{Introduction}
Insight into the three-dimensional (3-D) structure of proteins is fundamentally important to help us understand their cellular functions,
their roles in disease mechanisms, and for structure based development of pharmaceuticals. Recent advancements in cryogenic electron microscopy (cryo-EM), including better detector technologies and data processing techniques, have enabled high-resolution imaging of proteins and large biological complexes at the atomic scale \cite{callaway2020}.

To construct a structural, atomically detailed model for a protein, typically tens of thousands of single-particle images are collected, sorted and aligned to reconstruct a 3-D density map volume. 
Next, atomic coordinates are built into the density map. 
The latter routine is known as the map-to-model process, which typically requires a considerable amount of human intervention and inspection, notwithstanding the availability of automated tools to aid the process \cite{demaio2016, terwilliger2018a, terwilliger2018b}. 

Despite significant progress in machine learning techniques in 2-D or 3-D object detection \cite{huang2017, he2018, bochkovskiy2020, carion2020} and protein folding \cite{senior2020}, deep learning approaches to modeling atomic coordinates into cryo-EM densities remain relatively unexplored. 
Multiple research groups have proposed convolutional neural networks (CNNs) for detecting amino acid residues in a cryo-EM density map, but either did not address the final map-to-model step \cite{li2016, rozanov2018, subramaniya2019, mostosi2020}, or use a conventional optimization algorithm to construct the final model (see Related Work). 
Conventional search algorithms have high time- and space-complexity, constituting a bottleneck for large protein complexes, and are unable to exploit rich structural information encoded in genetic information \cite{senior2020}.  

\begin{figure}[htbp]
\centerline{\includegraphics[scale=1]{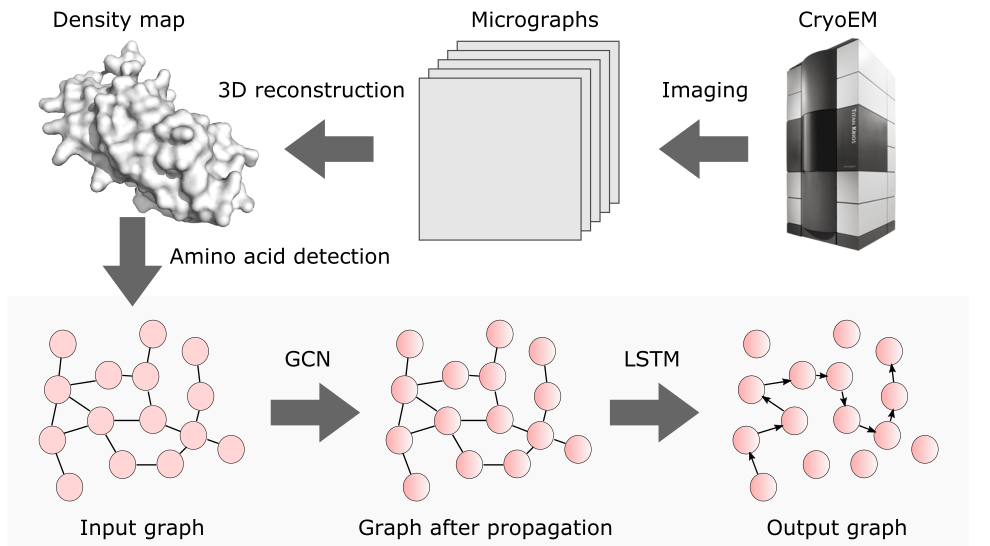}}
\caption{Overview of the map-to-model pipeline. The present work focuses on the bottom panel (shaded box), determining a structural model consistent with the protein sequence from candidate amino acid positions.}
\label{fig:intro}
\end{figure}

Here, we address these shortcomings by presenting an approach for protein structure determination from cryo-EM densities based entirely on neural networks. 
First, we use a 3D CNN with residual blocks \cite{he2015} we called RotamerNet to locate and predict amino acid and rotameric identities in the 3-D density map volume.
Next, we apply a graph convolutional network (GCN) \cite{kipf2017} to create a graph embedding using the nodes with 3D structural information generated by RotamerNet. 
Inspired by the UniRep approach \cite{alley2019}, we then apply a bidirectional long short-term memory (LSTM) module to select and impose an ordering consistent with the protein sequence on the candidate amino acids, effectively generating a refined version of the graph with directed edges connecting amino acids (Fig. \ref{fig:intro}). 
We trained our LSTM on sequence data (UniRef50) alone, taking advantage of structural information encoded in the vast amount of genetic information \cite{senior2020}. 
In this paper we focus on the protein structure generation part with a GCN and an LSTM, which together we termed the Structure Generator. Our main contributions are: 
\begin{itemize}
    \item The first, to our knowledge, entirely neural network based approach to generate a protein structure from a set of candidate 3D rotameric identities and positions.
    \item Exploitation of genetic information learned from UniRef50 sequences to help generate a 3-D structure from cryo-EM data using a GCN embedding and a bidirectional LSTM.
\end{itemize}

\section{Related work}

\subsection{Map-to-model for cryo-EM maps}
Over the last few years, cryo-EM has evolved as a major experimental technique for determining novel structures of large proteins and their complexes. Computational techniques to process and analyze the data, and build protein structures are challenged by this avalanche of data. For example, widely-used {\it de novo} cryo-EM structure determination tools, e.g., \verb+phenix.map_to_model+ \cite{terwilliger2018a, terwilliger2018b} or \verb+rosettaCM+ \cite{Song2013} partially automate cryo-EM data interpretation and reconstruction, but typically take many hours to generate a preliminary model and can require significant manual intervention. The underlying algorithms are often decades old, and are difficult to adapt to faster (e.g. graph processing unit, GPU) architectures. It will be critical to modernize these approaches, and capitalize on recent advances of deep learning and GPUs to expedite this procedure. 

Several deep-learning based approaches for protein structure determination from cryo-EM data have been proposed. Li and coworkers \cite{li2016} introduced a CNN-based approach to annotate the secondary structure elements in a density map, an approach later also proposed by Subramaniya \textit{et al.} and Mostosi \textit{et al.} \cite{subramaniya2019, mostosi2020}. The feasibility of an end-to-end map-to-model pipeline with deep learning has also been explored. Xu and colleagues trained a number of 3-D CNNs with simulated data to localize and identify amino acid residues in a density map and use a Monte-Carlo Tree Search (MCTS) algorithm to build the protein backbone \cite{xu2019}. Using an entirely different architecture, Si and coworkers divided the map-to-model procedure into several tasks addressed by a cascade of CNNs. However, their procedure also relied on a conventional Tabu-Search algorithm to produce the final protein model \cite{si2020}.

\subsection{Graph neural networks}
Graph neural networks \cite{scarselli2009, kipf2017} are natural representations for molecular structures with atoms as nodes and covalent bonds as edges. Duvenaud and coworkers pioneered  this approach using a GCN to learn molecular fingerprints, which are important in drug design \cite{duvenaud2015}. Numerous other applications of GCNs to predict or generate molecular properties can be found in the literature. For example, 
Li and colleagues demonstrated the utility of a generative GCN to construct 3-D molecules from SMILES strings, among other applications \cite{li2018}.

\subsection{Long short-term memory}
Recurrent neural networks (RNNs) are widely used in natural language processing tasks. Their  architecture is designed to process, classify, or predict properties of sequences as input and can output sequences with desired properties \cite{graves2013}.
Among many RNN architectures, the LSTM model was proposed to address the gradient vanishing problem for long sequences \cite{hochreiter1997} and a number of variants have since been studied to further increase its capacity, such as multi-layer and bidirectional LSTMs.
LSTMs are often used in conjunction with other neural network models. An image captioning system, for example, can be realized by using a 2-D CNN that extracts high-dimensional features from an image and an LSTM that outputs a sentence describing the input image \cite{donahue2015}. 

\section{Method}

\subsection{Model}

In this section we present the Structure Generator, a neural network model for protein model building consisting of a GCN and, subsequently, a bidirectional LSTM module. 
The input for the Structure Generator is a set of nodes labeled with 3-D coordinates and amino acid identity. 
To generate a set of candidate amino acids we have previously implemented RotamerNet (unpublished), a 3-D CNN based on the ResNet architecture \cite{he2015} that can identify amino acids and their rotameric identities in an EM map. This set of candidate amino acids is not constrained by the sequence, and and their 3-D locations are located based entirely on their density profiles. The set can contain false positives (an amino acid rotamer is proposed at a location where there is none) or false negative (a correct amino acid rotamer was not identified). RotamerNet outputs an amino acid and rotamer identity together with proposed coordinates for its C$\alpha$ atom. In the remainder, we will only consider the amino acid identity. A node $v$ is a proposed amino acid identity with the C$\alpha$ coordinates.

Next, we generate a C$\alpha$ contact map for all predicted C$\alpha$ coordinate locations. 
We connect any two proposed C$\alpha$ with a distance less than a given threshold ($4.0 \mathrm{\AA{}}$) with an undirected edge. 
We represent the input with two matrices: an $m$ by $20$ matrix of node features and an $m$ by $m$ adjacency matrix that describes the connectivity between nodes.

\begin{figure}[htbp]
\centerline{\includegraphics[scale=1]{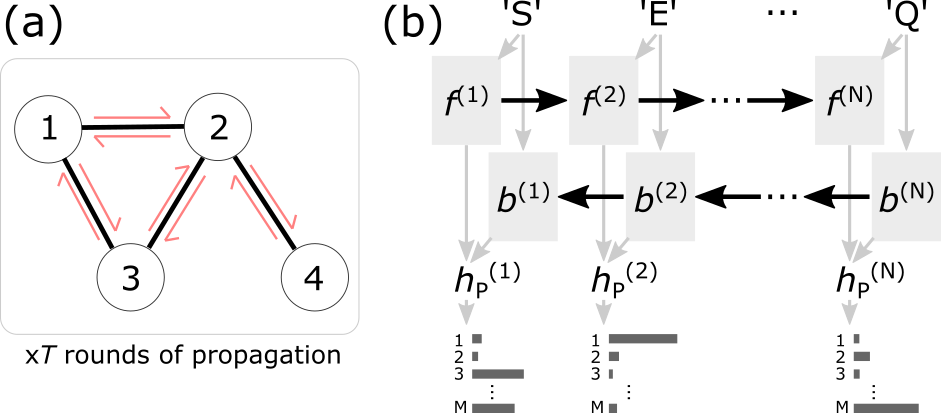}}
\caption{Architecture of the Structure Generator. (a) A graph convolutional network allows the embedding of each node to communicate through edges for $T$ rounds of propagation. (b) Protein sequence (SE...Q) is fed to the bidirectional LSTM to guide the modeling. The outputs from the forward ($f$) and backward ($b$) LSTM state at each time step are concatenated to predict the best node to add to the protein $P$. $N$ and $M$ are the length of the sequence and the number of nodes in the graph, respectively.}
\label{fig:archi}
\end{figure}

We generate a high dimensional embedding for each node $v$, $\mathbf{H}_{\mathrm{node}} = \mathrm{GCN}(\mathbf{A}, \mathbf{F})$, where $\mathbf{H} = [h_{\mathrm{node}}^{(1)\intercal}, \ldots, h_{\mathrm{node}}^{(m)\intercal}]^{\intercal}$, $\mathbf{A}$ is the adjacency matrix with $a_{i,j} = 1$ for each neighbor pair $(i, j)$ or the node itself, i.e. $i=j$, and $\mathbf{F} = [f_{\mathrm{node}}^{(1)\intercal}, \ldots, f_{\mathrm{node}}^{(m)\intercal}]^{\intercal}$ are input features. 
Features are generated with $f_{\mathrm{node}}^{(v)} = \mathrm{NN}(s^{(v})$, where $s \in \mathbb{R}^{20}$ is the normalized softmax score vector for a node $v$ obtained from the RotamerNet and $\mathrm{NN}\left(\cdot\right)$ denotes a single-layer neural network.
We implemented the GCN module following \cite{kipf2017} (Fig. \ref{fig:archi}(a)).
Note that the GCN can be applied in $T$ layers to propagate messages, thereby increasing the capacity of the network \cite{duvenaud2015, li2018}. 
As depicted in Fig. \ref{fig:archi}(a), in each of the GCN layer, messages propagate through edges, sharing the embedding of a node with its neighbors. 
For example when $T=2$, $\mathbf{H}_{\mathrm{node}} = \mathrm{GCN}^{(2)}(\mathbf{A}, \mathrm{GCN}^{(1)}(\mathbf{A}, \mathbf{F}))$, and these two GCN layers can share the same set or have different sets of parameters.  

The Structure Generator then uses a bidirectional LSTM module as a decoder for the refined protein chain generation. We use zero vectors for initial hidden and cell states.
At each time step $t$, an embedding of an amino acid in the sequence $h_{\mathrm{seq}}^{(t)} = \mathrm{NN}\left(c_{\mathrm{seq}}^{(t)}\right)$, where $c_{\mathrm{seq}}^{(t)} \in \mathbb{R}^{20}$ is a one-hot encoding for an amino acid at position $t$ in the sequence, is fed into the LSTM cell. 
The cell output at each time step, $h_{P}^{(t)}$, can be viewed as the current graph representation for $P$, the protein structure to be built.
A score $z_{\mathrm{node}}^{(v)} \in \mathbb{R}$ of a candidate node to be selected as the next node to be added to $P$  is determined by $z_{\mathrm{node}}^{(v)} = \mathrm{NN}\left(h_{\mathrm{add}}^{(v)} + h_{P}\right)$, where $h_{\mathrm{add}}^{(v)} = \mathrm{NN}\left(h^{(v)}\right)$.
At each time step $t$, the node with the highest softmax score $p_{\mathrm{node}}^{(v)} = \exp(z_{\mathrm{node}}^{(v)}) / \sum_{v'} \exp(z_{\mathrm{node}}^{(v')})$ is selected and added to $P$.
The selection process continues until the end of the sequence $t=N$, where $N$ is the length of the sequence, is reached, at which point the cross entropy loss is computed for the entire sequence in the training stage, or bitwise accuracy for the inference stage. 
We implemented the decoder with a bidirectional LSTM, in which the outputs from one LSTM fed with a forward sequence and another fed with a backward sequence are concatenated to obtain $h_{P}^{(t)}$ for each time step $t$. 
We found that a bidirectional LSTM consistently outperformed a uni-directional LSTM.
We also found that using the ensemble of inference results with a forward (from N-terminus) and a backward (from C-terminus) sequence further improves the accuracy.
Fig. \ref{fig:archi}(b) illustrate the generation process with the sequence as the input at each time step (top) and the best corresponding node prediction as output (bottom).
Importantly, the sequence information is used both in the training and inference stages to guide the protein modeling. 

\subsection{Training data}

To train the Structure Generator, we randomly selected 1,000,000 and 100 sequences with length in $[50, 450]$ from the UniRef50 dataset \cite{suzek2015} for the training and validation sets, respectively.
The remaining sequences (approximately 30 million) are kept untouched for future uses. 
The median and mean sequence length in the validation set are $174$ and $201.3$.

RotamerNet was trained on simulated density profiles of proteins, generated as follows. 
We selected high-quality protein structures from the Protein Data Bank, with resolution between $1.4$ and $1.8 \mathrm{\AA{}}$. We used \verb+phenix.fmodel+ to generate electron scattering factors with $10\%$ noise to simulate the cryo-EM density maps for $18,893$ protein structures, $98\%$ of which were used to train the RotamerNet. 

The orders of amino acid residues in a given protein structure are shuffled. 
Because the UniRef50 dataset has only protein sequences, we assumed perfect C$\alpha$-C$\alpha$ contact maps and simulated input features, i.e. a normalized softmax score vector $s \in \mathbb{R}^{20}$ by $s_i = |\mathcal{N} \sim (0, 0.01)|$ for $i \neq j$ and $s_j = 1 - \sum_{i \neq j} s_i$, where $j$ is the index corresponding the ground-truth amino acid identity. 
The ground truth sequence is used in both training and inference stages. 

\subsection{Training the Structure Generator}

We trained the Structure Generator with the ADAM optimizer with batch size $1$ and learning rate $0.001$ for first 100,000 iterations and decreased the learning rate to $0.0001$ for the rest.
The sum of the cross entropy loss at each sequence position with the ground true index $j\in \mathbb{R}$ and the vector of normalized scores $p \in \mathbb{R}^{m}$,
\begin{equation}
    \mathrm{Loss} = -\sum_{t=1}^{n} \log p_{j}
\end{equation}
where $n$ is the length of the ground truth sequence and $m$ is the number of nodes in the raw graph, is calculated and back-propagated through the entire network, i.e. the LSTM and then the GCN.
In the inference stage, the average accuracy
\begin{equation}
\mathrm{AA} = \frac{1}{K}\sum_{\mathrm{prot}} \frac{1}{N_{\mathrm{prot}}}\sum_{t=1}^{N_\mathrm{prot}} \mathbbm{1}(\hat{j}_t = j_t),
\end{equation}
i.e., the fraction of amino acids whose identity was predicted correctly, is used to evaluate the performance of the Structure Generator on a set of $K$ protein structures, where $N_{\mathrm{prot}}$ denotes the sequence length of a protein, $j_t$ and $\hat{j}_t$ are the ground truth and predicted note index for the $t$-th step in the LSTM, respectively.

We trained the GCN with sequence embedding dimension $32$, node embedding dimension $128$ and LSTM hidden state dimension $2 \times 256$ ($256$ for each direction).
During training, we added $(500 - n)$ dummy (false positive) nodes with random edges in each iteration to complicate the training data. 

\section{Results}

\begin{figure}[htbp]
\centerline{\includegraphics[scale=1]{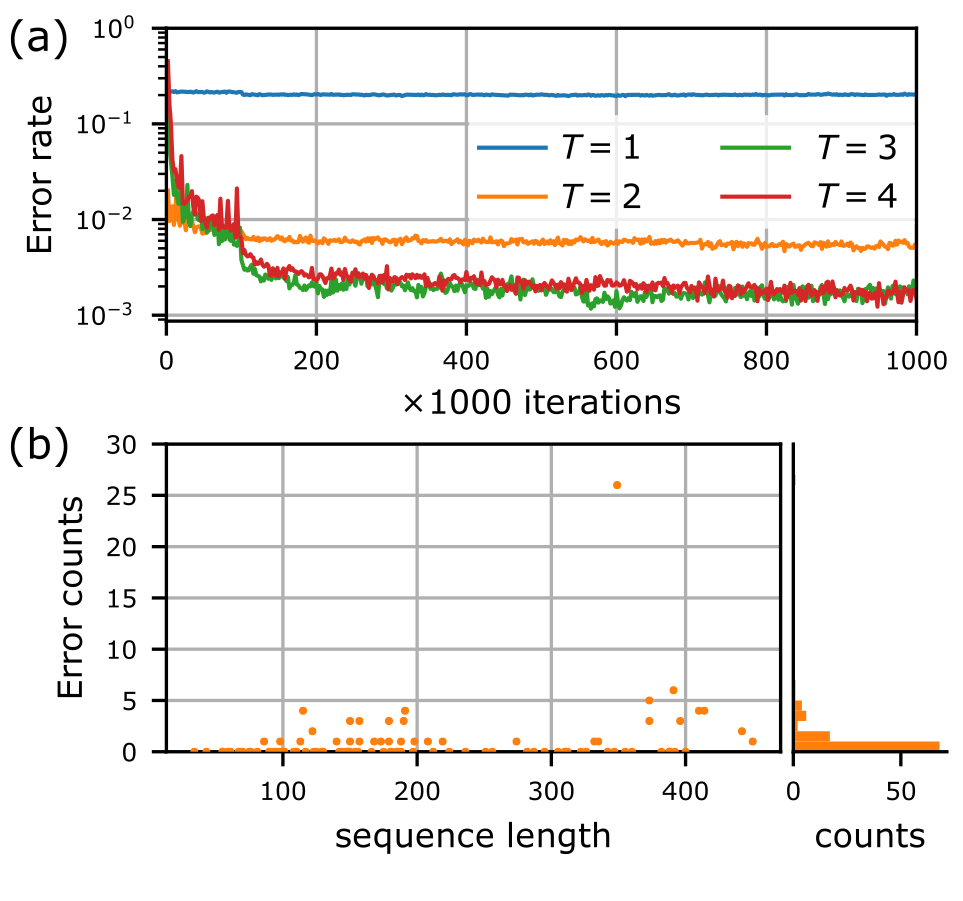}}
\caption{Validation results. $N=100$. (a) Error rate, defined as $1 - \textrm{average accuracy}$, vs. training iterations with GCNs with different layers. (b) Error counts, i.e. the number of incorrect amino acid assignments in a protein structure as a function of sequence length for the $T=2$ model.}
\label{fig:training}
\end{figure}

We first examined the effect of the GCN on structure determination. We found that the number of GCN layers can dramatically improve the average accuracy on the validation set. For example, using two rather than one GCN layer, i.e.  $T=1$ to $T=2$, yields an $20\%$ improvement (Fig. \ref{fig:training}(a)). 
Encouraged by this improvement, we further trained the Structure Generator with $T=\{3,4\}$.
Fig. \ref{fig:training}(a) shows the error rate ($1 - \textrm{AA}$) curves on the validation data for different number of GCN layers $T$.
While increasing to $T=3$ again gives another $0.3\%$ increase in average accuracy, $T=4$ adds only $0.04\%$ (Table \ref{table:ensemble}). 
This observation suggests that $T=2$, which can be interpreted as learning 5-mer spatial motifs in the graph (see discussion in \ref{sec:rotamernet}) is sufficient for the model to capture the implicit structural information in the graph and the sequence.
In the remainder, unless stated otherwise we fixed $T=2$ for inference in all experiments. 
Fig. \ref{fig:training}(b) shows the error counts on the 100 protein structure in the validation set as a function of sequence length, suggesting that the error counts increased very mildly with the length of sequence. 

To study the efficacy of the GCN, we also tested a GCN with $T=0$, i.e., the node features $\mathbf{F}$ are added directly to the LSTM outputs. 
This model dramatically reduced the average accuracy to $0.0347$, which is approximately the probability of randomly selecting the correct node out of candidate nodes that have the amino acid identity matching the sequence input.
This result indicates that a GCN embedding of RotamerNet's output is required  for the LSTM to predict an ordered graph consistent with the sequence and the GCN.


\begin{table}[]
\centering
\caption{\label{table:ensemble} Average accuracy on various validation datasets with different GCN layer and inference settings.}
\begin{tabular}{rccrrrr}
Dataset                                         & Inference & \multicolumn{1}{r}{$T=0$}  & $T=1$  & $T=2$  & $T=3$  & $T=4$  \\ 
\hhline{=======}
\multirow{2}{*}{UniRef50}                       & Forward   & \multicolumn{1}{r}{0.0347} & 0.7952 & 0.9949 & 0.9981 & 0.9985 \\
                                                & Ensemble  & \multicolumn{1}{r}{0.0347} & 0.8383 & 0.9954 & 0.9986 & 0.9987 \\ \hline
\multirow{2}{*}{ProteinNet}                     & Forward   & -                          & 0.8214 & 0.9899 & 0.9957 & 0.9960 \\
                                                & Ensemble  & -                          & 0.8577 & 0.9908 & 0.9960 & 0.9963 \\ \hline
\multicolumn{1}{l}{\multirow{2}{*}{RotamerNet}} & Forward   & -                          & 0.6162 & 0.7443 & 0.6912 & 0.6853 \\
\multicolumn{1}{l}{}                            & Ensemble  & -                          & 0.6409 & 0.7538 & 0.7060 & 0.6965
\end{tabular}
\end{table}

\subsection{Performance of the Structure Generator on the ProteinNet dataset}

Next, we evaluated the Structure Generator on the ProteinNet data set, a standardized machine learning sequence-structure dataset with standardized splits for the protein structure prediction and design community \cite{alquraishi2019}. The CASP12 ProteinNet validation set used here has 224 structures with sequence lengths ranging from 20 to 689, with median $163$ and mean $204.4$.
We selected the same parameters as those for the UniRef50 dataset to generate simulated feature vectors and generated C$\alpha$ contact maps based on the backbone atom coordinates from ProteinNet.
We note that a small number of C$\alpha$ coordinates are absent in ProteinNet owing to lack of experimental data. Compared to the UniRef50 validation set, the ProteinNet validation set is therefore more challenging, as edges in the input graph can be missing. 
Validation results on ProteinNet are given in the second row of Table \ref{table:ensemble}.

\begin{figure}[htbp]
\centerline{\includegraphics[scale=1]{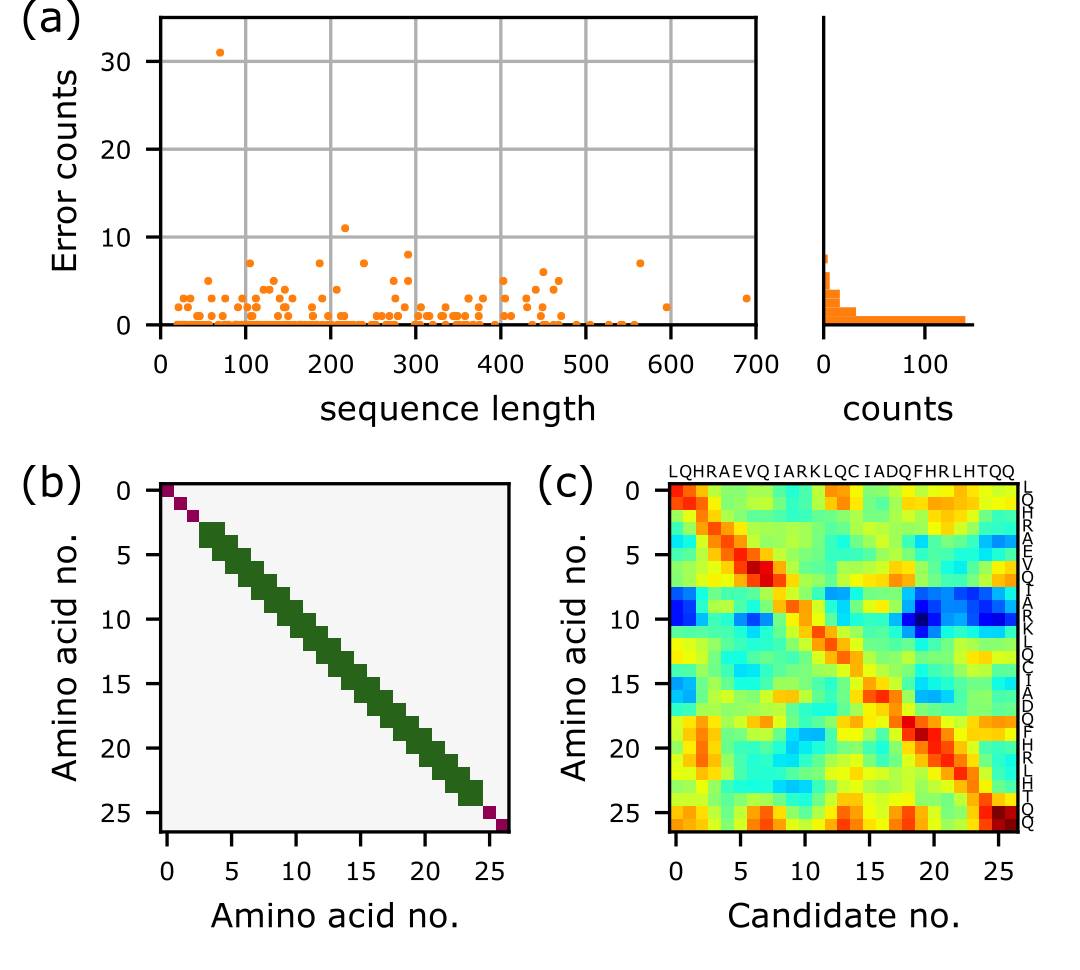}}
\caption{Test on the ProteinNet CASP12 validation set. (A) Error counts, i.e. number of incorrect amino acid  assignments in a protein structure as a function of sequence length, and the histogram, with the $T=2$ model. (b) Contact map of prosurvival protein A1 (PDB ID 2vog). Green pixels are contacts and purple pixels indicate where C$\alpha$ coordinates are unknown and thus the contacts are missing. (c) The output of the Structure Generator on 2vog. Red means higher probability whereas blue means less likely.}
\label{fig:proteinnet}
\end{figure}

Nonetheless, well over $50\%$ of the structures in the data set are correctly predicted without any errors (Fig. \ref{fig:proteinnet}(a)).
Remarkably, the Structure Generator can correctly predict amino acids for which the C$\alpha$ records are missing. For example, atomic coordinates for the first three and last two amino acids of prosurvival protein A1 (PDB ID 2vog) are missing, but the Structure Generator can still completely reconstruct the protein model (Fig. \ref{fig:proteinnet}(b)). 
Several amino acids, for example glutamine (Q), occur multiple times in the sequence.
As a result, the last two rows in the Structure Generator output have repeating patterns (Fig. \ref{fig:proteinnet}(c)), which did not prevent the Structure Generator from predicting the correct nodes for each of the positions corresponding to glutamine.

\subsection{Performance of the Structure Generator on RotamerNet data\label{sec:rotamernet}}

\begin{figure}[t]
\vspace{-4.5cm}
\centerline{\includegraphics[scale=1]{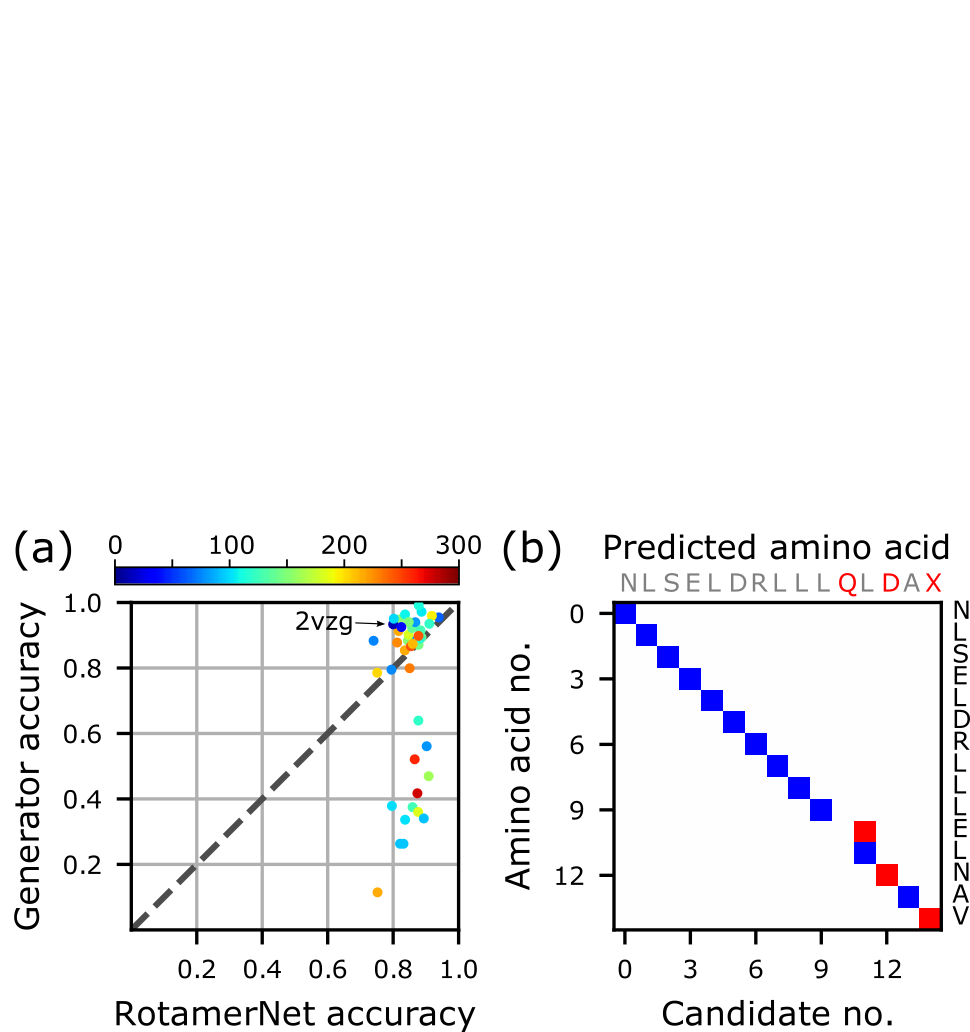}}
\caption{Results on the RotamerNet data. (a) Structure Generator accuracy vs. RotamerNet accuracy. Each data point represents a protein structure. The color code indicates the length of the structure. (b) Amino acid position of a select structure, PDB ID 2vzg, predicted by the Structure Generator. Red pixels are where RotamerNet made incorrect prediction on the amino acid type. Sequence on the top is derived from RotamerNet output and on the right is the ground truth.}
\label{fig:rotamernet}
\end{figure}

Finally, to demonstrate the utility of our approach with an upstream machine learning approach, we tested the Structure Generator on output from  RotamerNet. 
The RotamerNet validation data set consists of the amino acid type classification scores for simulated cryo-EM density maps from $45$ protein structures of various lengths from $15$ to $278$ with various nominal resolution ranging from $1.4$ to 1.8$\mathrm{\AA{}}$.
The average accuracy of the RotamerNet validation set is $0.852$, meaning that a non-trivial fraction of input features for the Structure Generator is noisy or incorrect.
Fig. \ref{fig:rotamernet}(a) shows the RotamerNet and the Structure Generator accuracy of the $45$ proteins with various sequence lengths.
Remarkably, while the performance of the Structure Generator is largely limited by the RotamerNet accuracy (data points beneath the dashed gray line in Fig. \ref{fig:rotamernet}(a)), as indicated by the correlation, a number of proteins have higher Structure Generator accuracy than RotamerNet accuracy, suggesting that the Structure Generator can tolerate and recover errors from upstream machine learning approaches.
To understand the characteristics of the Structure Generator, we plot the confusion matrix for the C-terminal calponin homology domain of alpha-parvin (PDB ID 2vzg).
Among those amino acids whose identity and position are correctly predicted by RotamerNet and the Structure Generator (Fig. \ref{fig:rotamernet}(b), blue dots on the diagonal), there are two red dots indicating that the prediction error from RotamerNet does not necessarily prevent the Structure Generator from making correct predictions.  
Again we point out that the Structure Generator has been trained only on UniRef50 sequences and simulated features, and has not been fine-tuned with the RotamerNet data. 
We anticipate that either doing so or training with the upstream model will further enhance the performance of the Structure Generator.   

We observe in Table \ref{table:ensemble} that the $T=2$ model performs best on the RotamerNet data. The Structure Generator relies on learning the correlation between the graph embedding and the motifs in the sequence. 
Increasing the number of GCN layers in principle allows the Structure Generator to recognize longer n-grams and spatial motifs of increased connectivity length. However, such correspondences will become increasingly noisy as lengths increase. Based on this observation, we therefore suggest that $T=2$ is a practical choice.

\section{Conclusion}
Building an atomic model into a map is a time- and labor-intensive step in single particle cryo-EM structure determination, and mostly relies on traditional search algorithms that cannot exploit recent advancements in GPU computing and deep learning. 
 To address these shortcomings, we have presented the Structure Generator, a full-neural network pipeline that can build a protein structural model from a set of unordered candidate amino acids generated by other machine learning models. 
Our experiments show that a GCN can effectively encode the output from the upstream model as a graph while a bidirectional LSTM can precisely decode and generate a directed amino acid chain, even when the input contains false or erroneous entries.
Our experiments suggest that a two-layer GCN is sufficient for processing the raw graph while preventing over-fitting to the training data. 

The Structure Generator exploits genetic information to guide the protein structure generation, and showed promising results on the RotamerNet data set without fine tuning. Training on the ProteinNet dataset and fine-tuning on the RotamerNet dataset will further enhance performance.
While a practical machine learning model for cryo-EM map-to-model is still a work-in-progress, in part because of the lack of high-resolution experimental data \cite{nakane2020} for training, our proposed framework can complement the existing approaches and ultimately pave ways toward a fully trainable end-to-end machine learning map-to-model pipeline, making human intervention-free protein modelling in a fraction of a minute possible. 

\section*{Current affiliation}
This work was initiated when S.H.d.O. and H.v.d.B. were at SLAC National Accelerator Laboratory.
S.H.d.O. is currently at Frontier Medicines, CA, USA.
In addition to his position at Atomwise, H.v.d.B. is on the faculty of the Department of Bioengineering and Therapeutic Sciences, University of California, San Francisco, CA, USA.

\end{document}